\documentclass[oldversion]{aa}

\usepackage{graphicx}
\usepackage{txfonts}

\usepackage{natbib}
\bibpunct{(}{)}{;}{a}{}{,}

\begin{document}

\title{Low redshift AGN in the Hamburg/ESO Survey: I. The local AGN luminosity function
       \thanks{Based on observations collected at the European Southern Observatory, 
               Chile (Proposal 145.B-0009).}
       }
 
\author{Andreas Schulze \and Lutz Wisotzki \and Bernd Husemann}
\institute{Astrophysikalisches Institut Potsdam, An der Sternwarte 16, 
	14482 Potsdam, Germany }
\offprints{A. Schulze, \email{aschulze@aip.de} }

\date{Received 22 June 2009 / Accepted 17 September 2009}

\abstract{%
We present a determination of the local ($z\approx0$) luminosity function of optically selected type~1 (broad-line) Active Galactic Nuclei. Our primary resource is the Hamburg/ESO Survey (HES), which provides a well-defined sample of more than 300 optically bright AGN with redshifts $z<0.3$ and blue magnitudes $B \la 17.5$. AGN luminosities were estimated in two ways, always taking care to minimise photometric biases due to host galaxy light contamination. Firstly, we measured broad-band $B_J$ (blue) magnitudes of the objects over small apertures of the size of the seeing disk. Secondly, we extracted H$\alpha$ and H$\beta$ broad emission line luminosities from the spectra which should be entirely free of any starlight contribution. Within the luminosity range covered by the HES ($-19\ga M_{B_J}\ga -26$), the two measures are tightly correlated. The resulting AGN luminosity function (AGNLF) is consistent with a single power law, also when considering the effects of number density evolution within the narrow redshift range. We compared our AGNLF with the H$\alpha$ luminosity function of lower luminosity Seyfert~1 galaxies by Hao et al. (2005) and found a smooth transition between both, with excellent agreement in the overlapping region. From the combination of HES and SDSS samples we constructed a single local AGNLF spanning more than 4 orders of magnitude in luminosity. It shows only mild curvature which can be well described as a double power law with slope indices of $-2.0$ for the faint end and $-2.8$ for the bright end. We predicted the local AGNLF in the soft X-ray domain and compared this to recent literature data. The quality of the match depends strongly on the adopted translation of optical to X-ray luminosities and is best for an approximately constant optical/X-ray ratio. We also compared the local AGNLF with results obtained at higher redshifts and find strong evidence for luminosity-dependent evolution, in the sense that AGN with luminosities around $M_B \simeq -19$ are as common in the local universe as they were at $z=1.5$. This supports the 'AGN downsizing' picture first found from X-ray selected AGN samples.
}

\titlerunning{The local AGN luminosity function}
\authorrunning{Schulze et al.}

\maketitle

\section{Introduction}
A good knowledge of the AGN luminosity function (AGNLF) is important for our understanding of the AGN population and its evolution, as well as for gaining insight into the history of black hole growth and galaxy evolution \citep[e.g.][]{Yu:2002,Marconi:2004}. Thanks to recent heroic quasar surveys such as the 2dF QSO Redshift Survey (2QZ) and the Sloan Digital Sky Survey (SDSS), large samples of AGN became available, and the optical AGN/QSO luminosity function is today well established over a wide range in redshifts ($0.3\la z  \la 5$) and luminosities \citep{Boyle:2000,Croom:2004,Richards:2006a,Bongiorno:2007,Croom:2009}. 
However, neither of these surveys reaches below redshifts of $z\sim 0.3$--0.4, because the imposed colour selection criteria, and also the discrimination against extended sources in the 2QZ sample, exclude a large fraction of lower luminosity, `Seyfert-type' AGN. 
The \emph{local} AGNLF is therefore much less constrained than at higher redshifts, despite of its importance as a zero-point for studies on the AGNLF and quasar evolution.

Approaching this problem from the other end, large spectroscopic galaxy surveys are a powerful way to unravel the AGN content of local galaxy samples \citep[e.g.][]{Huchra:1992,Ulvestad:2001,Hao:2005}. When constructing an AGNLF from galaxy surveys, care has to be taken in defining `AGN luminosity': Simply taking the optical galaxy magnitude will lead to an ill-defined mix of host galaxy and AGN contributions. It is much better to base the AGNLF on the conspicuous broad Balmer emission lines which avoids the host galaxy contamination problem. This approach was recently adopted by \citet{Hao:2005} and \citet{Greene:2007a}. But since AGN are scarce among galaxies unless the selection is sensitive to very low levels of nuclear activity \citep{Ho:1997}, any general galaxy survey has a low yield, and the samples are dominated by weak Seyfert nuclei and rarely reach quasar-like luminosities.

In this paper we study the population statistics of low-redshift quasars and moderately luminous Seyferts. We exploit the Hamburg/ESO Survey \citep[HES; e.g.][]{Wisotzki:2000c}, a survey which was specifically designed to address the selection of low-redshift AGN with visible host galaxies, and to provide a well-defined, complete, and unbiased sample of the local AGN population. Here, the term `complete' is meant in a methodological sense, implying that \emph{all} AGN selected by the criteria are included in the sample. Since obscured or heavily reddened AGN are typically not found in the HES, our sample can only claim to be representative of the unobscured `type~1' AGN with broad emission lines in their spectra. We refrain in this paper from repeatedly recalling this fact, but it should be understood that we always imply this restriction when using the term `AGN'.

This is the third paper investigating the local AGNLF in the HES.  \citet{Koehler:1997} used a small sample of 27 objects obtained during the initial period of the survey to construct the first published local LF of quasars and Seyfert~1 galaxies.  \citet{Wisotzki:2000a} exploited $\sim 50$~\% of the HES to study the evolution of the quasar luminosity function and provided a re-determination of the local LF as a by-product. With the completed HES covering almost 7000~deg$^2$ of effective area, the AGN samples are doubled in size, as well as augmented by spectroscopic material. Thus a new effort is well justified. In the present paper we go substantially beyond our previous work not only in quantitative, but also in qualitative terms. We first present our data material and our treatment of the spectra. After constructing the standard broad-band AGNLF we investigate the H$\alpha$ and H$\beta$ emission line luminosity functions. This enables us to compare and combine our data with the recent work by \citet{Hao:2005} based on the SDSS galaxy sample, to obtain a local AGNLF covering several orders of magnitude in luminosities. We discuss our results in the context of other surveys and luminosity functions.

Thoughout this paper we assume a concordance cosmology, with a Hubble constant of $H_0 = 70$ km s$^{-1}$ Mpc$^{-1}$, and cosmological density parameters $\Omega_\mathrm{m} = 0.3$ and $\Omega_\Lambda = 0.7$ \citep{Spergel:2003}.

\section{Data   \label{sec:data}}

\subsection{The Hamburg/ESO Survey    \label{sec:hes}}

The Hamburg/ESO Survey (HES) is a wide-angle survey for bright QSOs and other rare objects in the southern hemisphere, utilising photographic objective prism plates taken with the ESO 1~m Schmidt telescope on La Silla. Plates in 380 different fields were obtained and subsequently digitised at Hamburg, followed by a fully automated extraction of the slitless spectra. In total, the HES covers a formal area of $\sim$\,$9500$~deg$^2$ in the sky. For each of the $\sim$\,$10^7$ objects extracted, spectral information at 10--15~\AA\ resolution in the range $3200\: \mathrm{\AA}\la\lambda\la 5200\:\mathrm{\AA}$ is recorded. Details about the survey procedure are provided by \citet[]{Wisotzki:2000c}.

The relatively rich information content in the HES slitless spectra enabled us to apply a multitude of selection criteria, depending on the object type in question. AGN can be easily recognised from their peculiar spectral energy distributions if they contain a sufficiently prominent nonstellar nucleus, i.e.\ `type~1' AGN. The HES picks up quasars with $B \la 17.5$ at redshifts of up to $z\simeq 3.2$. Several precautions ensured that low-redshift, low-luminosity AGN are not systematically missed. For example, both the extraction of spectra and the criteria to select AGN candidates contained a special treatment of extended sources, making the selection less sensitive to the masking of AGN by their host galaxies. This property makes the HES unique among optical survey in that it targets almost the entire local (low-redshift) population of type~1 AGN, from the most luminous quasars to relatively feeble low-luminosity Seyfert galaxies. It is this property which we exploit in the present paper.

\subsection{Photometry     \label{sec:phot}}

Photometric calibration in the HES is a two-step process. We first measured internal
isophotal broad-band $B_J$%
\footnote{$B_J$ is a roughly rectangular passband defined by a red cutoff at 5400~\AA\  (as imposed by the blue-sensitive photographic emulsion Kodak IIIa-J) and a blue cutoff at 3950~\AA. For AGN-like spectra, $B \approx B_J + 0.1$, in the Vega system.} 
magnitudes in the \emph{Digitized Sky Survey} (DSS) and calibrated these against external CCD photometric sequences. At the start of the HES around 1990, only the \textit{Guide Star Photometric Catalog} \citep[GSPC][]{Lasker:1988} was available which typically provided standard stars with $8 \la B \la 14$, clearly insufficient to obtain reasonably accurate photometry near $B\sim 17\dots 18$, the faint limit of the HES. We therefore launched a campaign to obtain our own deeper photometric CCD sequences in as many fields as possible; the results of this campaign will soon be made public (Schulze et al., in preparation). Augmenting these sequences by incorporating also the \textit{Guide Star Photometric Catalog} II \citep[GSPC-II;][]{Bucciarelli:2001}, we arrived at a satisfactory photometric calibration of all 380 HES fields.

However, these isophotal DSS magnitudes suffer from a number of drawbacks, the most undesirable one (for our purposes) being the fact that for extended objects, the nuclear AGN contribution tends to be drowned out by the host galaxy. Using isophotal magnitudes would have a detrimental effect on the estimation of a proper AGN luminosity function especially at the low-luminosity end. Other negative properties of the DSS magnitudes include the effects of relatively poor seeing in many fields, boosting the number of undesired blends.

In order to approximate the concept of \emph{nuclear magnitudes}, we optimised the extraction procedure of slitless spectra in the HES digitised data to assume either true point sources or point sources embedded in a diffuse envelope. In other words, the spectra used for the HES candidate selection always refer to an area of the size of the central seeing disk only (note also that the HES spectral plates were typically obtained under much better seeing conditions than the DSS). We then measured `nuclear magnitudes' by integrating the spectra over the $B_J$ passband, and calibrated these magnitudes against the DSS using hundreds of stars in each field. We demonstrate below (\S~\ref{sec:lummag}) that this procedure indeed produces measurements that in reasonably good approximation can be taken as basis for nuclear luminosities (see also Figs.~\ref{fig:mag_comp} and \ref{fig:luminosities} below).

An additional advantage of our approach lies in the fact that we thus entirely avoid variability bias: Selection criteria and fluxes in each field are all defined for a single dataset.

\begin{figure}
\resizebox{\hsize}{!}{\includegraphics*[angle=-90]{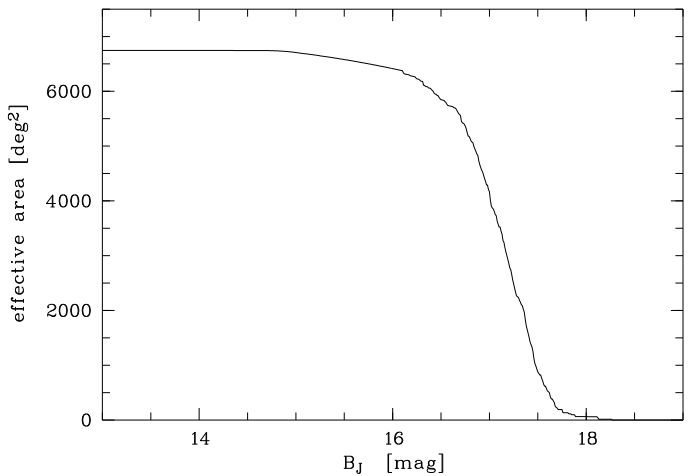}}
\caption{Effective area of the Hamburg/ESO survey as a function of the Galactic extinction corrected $B_J$ magnitude.}
\label{fig:effA}
\end{figure}

For the purpose of the present paper, a good knowledge of the survey selection function, and thus of the accurate flux limits is important. In the case of the HES, the
flux limit varies considerably between individual fields, mostly due to changes in the seeing and night-sky conditions, but also because of plate quality. Thus, each field maintains its own flux limit, always corresponding to the same limiting signal to noise ratio in the slitless spectra. As long as redshift-dependent selection effects are neglected, the survey selection function is identical to the `effective area' of the survey, combining the 380 flux limits into single array that provides the total survey area as a function of magnitude. The effective area of the entire HES is shown in Fig.~\ref{fig:effA}. Notice that this array also fully accounts for the losses due to overlapping spectra and images (see the full discussion in \citealt[]{Wisotzki:2000c}). 

All magnitudes used in this paper have been corrected for Galactic extinction, using the dust maps of \citet{Schlegel:1998} (averaged over each HES field), and the extinction law of \citet{Cardelli:1989}. Note that for the HES we had previously followed a different extinction correction recipe based on measured column densities of Galactic neutral hydrogen converted into extinction \citep[see][]{Wisotzki:2000c}. The main result of this change is a slight decrease of the average adopted extinction along most lines of sight.

Figure~\ref{fig:effA} shows that the average extinction-corrected limiting magnitude of the HES is $B_J < 17.3$, but with a dispersion of 0.5~mag between individual fields. There is essentially no bright limit, with 3C~273 recovered as the brightest AGN in the sample.

\subsection{Spectroscopic data      \label{sec:spec}}

All AGN and QSO candidates brighter than the flux limit in a given field were subjected to follow-up spectroscopy for confirmation and accurate redshift determination, altogether approximately 2000 targets. These observations were carried out during 23 observing campaigns between 1990 and 2000, using various telescopes and instruments at the ESO La Silla observatory. Known AGN that were recovered by the HES selection criteria were initially not part of the follow-up scheme. However, such objects were often included as backup targets, so that our spectroscopic coverage is almost complete for the entire AGN sample. For some quasars in areas overlapping with the Sloan Digital Sky Survey we could later add spectra from the public SDSS database.

Because of the diversity of telescopes used, the quality of the spectra varied significantly,  in signal to noise as well as spectral resolution.  We estimated the latter quantity from the width of the night sky \ion{O}{i} emission line at 5577~\AA. A list of individual observing campaigns with their estimated spectral resolution and the number of AGN from our sample observed is listed in Table~\ref{tab:camp}. 

Although a formal spectrophotometric calibration was available for all campaigns, the combined dataset is much too heterogeneous to allow for any consistent direct measurement of emission line fluxes. We therefore adjusted all spectra to a common, homogeneous flux scale by computing a synthetic $B_J$ magnitude from each spectrum and matching this to the `nuclear magnitude' photometry of the HES. This step again also ensured that AGN variability is of no importance.

\begin{table}
\caption{Observing campaigns for follow-up spectroscopy}
\label{tab:camp}
\begin{center}
\begin{tabular}{ccrr}
\hline \hline \noalign{\smallskip} 
   Date & Telescope & 
   \multicolumn{1}{c}{No. of Spectra} & 
   \multicolumn{1}{c}{Resolution $\lambda / \Delta\lambda$} \\ 
\noalign{\smallskip} \hline \noalign{\smallskip}
Dec 1990 & 3.6 m & 4 & 1020 \\ 
Apr 1991 & 3.6 m & 1 & 830 \\ 
Feb 1992 & 2.2 m & 4 & 370 \\ 
Feb 1992 & 3.6 m & 6 & 900 \\ 
Sep 1992 & 3.6 m & 4 & 700 \\ 
Mar 1993 & 3.6 m & 6 & 770 \\ 
Feb 1994 & 1.5 m & 1 & 1900 \\ 
Sep 1994 & 1.5 m & 2 & 1980 \\ 
Sep 1994 & 3.6 m & 4 & 680 \\ 
Nov 1994 & 1.5 m & 11 & 770 \\ 
Oct 1995 & 1.5 m & 30 & 650 \\ 
Oct 1995 & 2.2 m & 1 & 300 \\ 
Mar 1996 & 1.5 m & 19 & 590 \\ 
Oct 1996 & 1.5 m & 8 & 580 \\ 
Feb 1997 & 1.5 m & 24 & 600 \\ 
Oct 1997 & 1.5 m & 19 & 540 \\ 
Dec 1997 & 1.5 m & 33 & 540 \\ 
Sep 1998 & 1.5 m & 26 & 670 \\ 
Nov 1998 & 1.5 m & 23 & 600 \\ 
Sep 1999 & 1.5 m & 28 & 580 \\ 
Mar 2000 & 1.5 m & 15 & 540 \\ 
Sep 2000 & 1.5 m & 36 & 640 \\ 
Nov 2000 & 1.5 m & 22 & 790 \\ 
\noalign{\smallskip} \hline
\end{tabular}
\end{center}
\end{table}

\subsection{The sample     \label{sec:sample}}

For the present investigation of the `local' AGN population we selected all AGN from the final HES catalogue (Wisotzki et al., in prep.) that belong to the `complete sample' and that are located at redshifts $z < 0.3$. This sample contains 329 type-1 AGN. For 5 of them we could not obtain spectra of sufficient quality. Thus our sample is $324/329 \approx 98.5$~\% complete in terms of spectroscopic coverage.

\begin{figure}
\resizebox{\hsize}{!}{\includegraphics*[angle=-90]{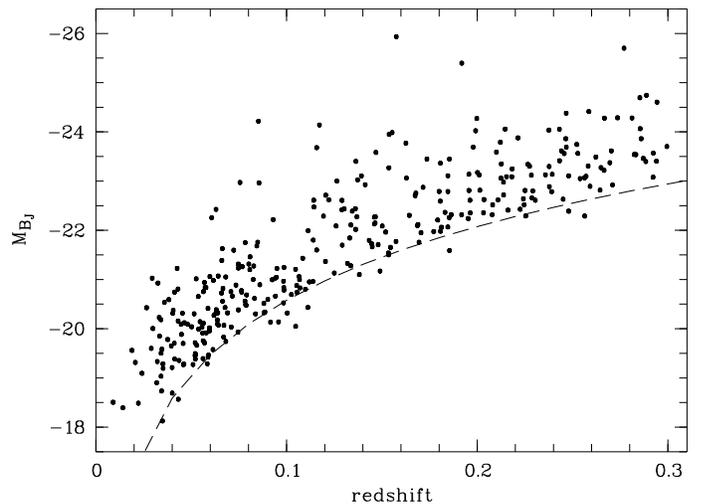}}
\caption{Redshift against absolute magnitude $M_{B_J}$. The dashed line indicates a constant apparent magnitude of 17.5~mag, approximately the flux limit of the HES. }
\label{fig:z_mag}
\end{figure}

The HES nuclear fluxes were converted into absolute $M_{B_J}$ magnitudes
using the $K$ correction of \citet[]{Wisotzki:2000b}.  Figure \ref{fig:z_mag} shows the distribution of the sample over redshifts and absolute magnitudes. A wide range of nuclear luminosities is covered, ranging from bright quasars with $M_{B_J} \simeq -26$ to low-luminosity Seyfert 1 galaxies of only $M_{B_J} \simeq -18$.

\section{Emission line properties      \label{sec:lineprop}}

\subsection{Fitting procedure      \label{sec:fit}}

As our sample is defined by $z < 0.3$, all objects have H$\beta$ and most also 
have H$\alpha$ visible in their spectra. We measured luminosities and line widths of the H$\alpha$ and H$\beta$ emission lines by fitting the spectral region around each line with a multi-component Gaussian model. Over this short wavelength range we approximated the underlying continuum as a straight line. While the structure of the broad lines is generally complicated, it has repeatedly been demonstrated \citep[e.g.][]{Steidel:1991,Ho:1997,Hao:2005b} that in many cases a double Gaussian provides an acceptable fit to the BLR lines. We always started with a single Gaussian per line. A second (in a few cases also a third) component was added only if a single Gaussian yielded a poor fit and the addition of a component significantly improved it.
At our limited spectral resolution, any possibly present narrow component of a broad line such as H$\beta$ is difficult to detect. We did not include an unresolved narrow component by default and added such a component only when it was clearly demanded by the data. 

The \ion{Fe}{ii} emission complex affecting the red wing of H$\beta$ can be sufficiently approximated by a double Gaussian at wavelength $\lambda\lambda$4924,\,5018 \AA\ for our data. We fixed their positions and  also fixed their intensity ratio $\lambda 5018 / \lambda 4924$ to 1.28, typical for BLR conditions. Thus only two parameters were allowed to vary, the amplitude and the line width. The [\ion{O}{iii}] $\lambda\lambda$4959,\,5007 \AA\, lines were also fitted by a double Gaussian with the intensity ratio fixed to the theoretical value of 2.98 \citep[e.g.][]{Dimitrijevic:2007}. The relative wavelengths of the doublet were fixed as well, but the position of the doublet relative to H$\beta$ was allowed to vary as a whole.

For the H$\alpha$ line complex, the contributions of [\ion{N}{ii}], and sometimes also [\ion{S}{ii}] needed to be taken into account. The latter was mostly well separated and could be ignored, but when required we modelled it as a double Gaussian with fixed positions and same width for both lines. For fitting the [\ion{N}{ii}] $\lambda\lambda$6548,\,6583 \AA\ doublet we left only the amplitude free. The positions were fixed, the intensity ratio was set to 2.96 \citep{Ho:1997}, and the line width was fixed to the width of the narrow [\ion{O}{iii}] lines. 

\begin{figure}
\centering
\resizebox{\hsize}{!}{\includegraphics*{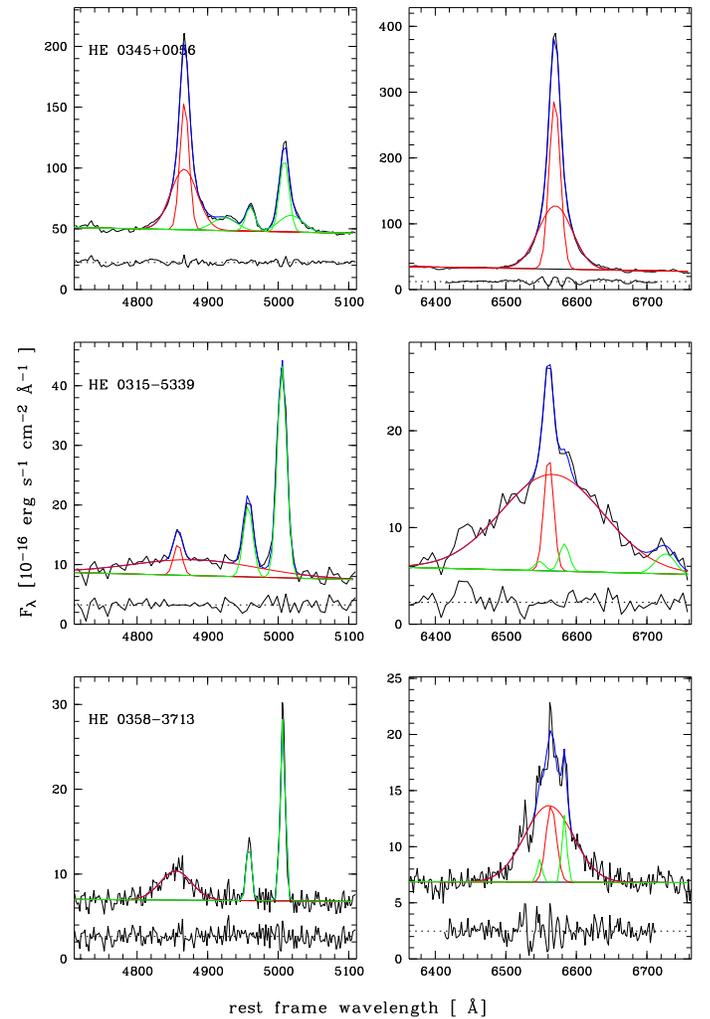}}
\caption{Examples of fits to the spectra, illustrating the quality of the spectroscopic data. The line complex of H$\beta$ is shown on the left side, the corresponding complex around H$\alpha$ is shown on the right side. The data are represented by black lines, the multi-component Gaussian fits to the Balmer lines are shown in red, other lines are shown green, and the combined model is overplotted in blue. Each panel also shows the fit residuals at the bottom.}
\label{fig:spec}
\end{figure}

With this set of constraints, each spectrum was fitted with a multi-Gaussian plus continuum model. We decided manually which model fits best, neglecting contributions by \ion{Fe}{ii}, [\ion{N}{ii}] and [\ion{S}{ii}] lines if not clearly present. In 6 objects we detected H$\alpha$ but the S/N was too low to trace H$\beta$. On the other hand, 21 of our spectra did not reach sufficiently into the red to cover H$\alpha$; further 8 spectra were too heavily contaminated  by telluric absorption to produce a reliable H$\alpha$ fit. For these objects, H$\beta$ was readily detected.

Of the 324 spectra of the sample, we thus could obtain reasonable fits for 318 objects in H$\beta$, and for 295 in H$\alpha$. Figure~\ref{fig:spec} shows some example results, illustrating the range of signal to noise ratios and resolutions of the spectral material.

\subsection{Line Fluxes    \label{sec:lfluxes}}

From the fitted model we determined the emission line fluxes of H$\alpha$ and H$\beta$ as well as the continuum flux at 5100 \AA.  As said above, a narrow component of the Balmer line was only subtracted if clearly identified in the fit. This happened in 46 cases for H$\alpha$ and in 34 instances for H$\beta$. We also measured the line widths of H$\alpha$ and H$\beta$, which were then used to estimate the black hole masses for the sample. These results are presented in a companion paper (Schulze \& Wisotzki, in prep.).

In order to estimate realistic errors we constructed artificial spectra for each object, using the fitted model and Gaussian random noise corresponding to the measured S/N. We used 500 realizations for each spectrum. We fitted these artificial spectra as described above, fitting the line and the continuum and measured the line widths and the line flux. The error of these properties was then simply taken as the dispersion between the various realizations. Note that this method provides only a formal error, taking into account fitting uncertainties caused by the noise. Other sources of error may include: A residual \ion{Fe}{ii} contribution; an intrinsic deviation of the line profile from our multi-Gaussian model, as well as uncertainties in the setting of the continuum level.

\begin{figure}
\resizebox{\hsize}{!}{\includegraphics*[angle=-90]{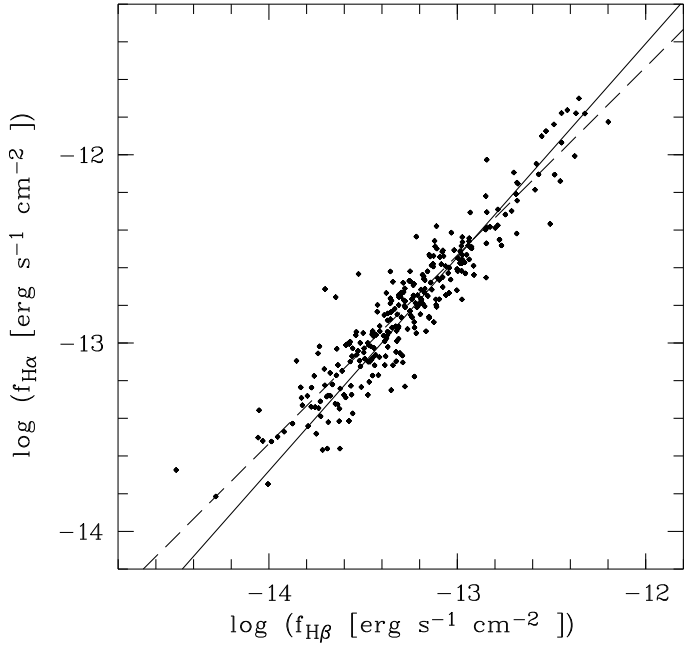}}
\caption{Correlation between the fluxes $f(\mathrm{H}\beta)$ and $f(\mathrm{H}\alpha)$. The solid line shows the regression result using the FITEXY method. The dashed line shows the best fit line with slope of unity.}
\label{fig:reg}
\end{figure}

\subsection{Relation between H$\beta$ and H$\alpha$ fluxes     \label{sec:relation}}

In order to investigate the distribution of AGN emission line luminosities we focus
on H$\alpha$ and H$\beta$ as the two most prominent recombination lines in our spectra. It is of interest to look at the relation between these two lines. While recent published work on this subject has mostly relied on H$\alpha$ \citep[][]{Hao:2005,Greene:2005}, extending similar studies to nonzero redshifts is easier using H$\beta$.  In Fig.~\ref{fig:reg} we plot the fluxes of the two broad lines against each other. As expected, H$\alpha$ and H$\beta$ are strongly correlated. To quantify this, we applied a linear regression between H$\alpha$ and H$\beta$ in logarithmic units, using the FITEXY method \citep{Press:1992} which allows for errors in both coordinates.  We account for intrinsic scatter in the relation following \citet{Tremaine:2002} by increasing the uncertainties until a $\chi^2$ per degree of freedom of unity is obtained. The best-fit relation found for the line fluxes is
\begin{equation}
	\log\left(f_{\mathrm{H}\alpha} \right) 
	=  (1.14 \pm 0.02)\, \log \left( f_{\mathrm{H}\beta} \right)  +  (0.46 \pm 0.01) 
	\label{eq:reg_flux}
\end{equation} 
where the fluxes are given in $10^{-13}$ erg s$^{-1}$ cm$^{-2}$. The rms scatter around the best fit is 0.15~dex. The relation between H$\alpha$ and H$\beta$ line flux using the FITEXY method is shown  as the solid line in Fig.~\ref{fig:reg}. However, a relation with a slope of unity is also consistent with the data (with a scatter of 0.14~dex).  When fixing the slope to one, the normalisation corresponds to the relation $f_\mathrm{H\alpha} = 2.96\, f_\mathrm{H\beta}$, consistent with the Case~B recombination value of 3.1 \citep[e.g.][]{Osterbrock:1989}. We conclude that the measured H$\alpha$ and H$\beta$ broad line fluxes give consistent results.

\begin{figure}
\resizebox{\hsize}{!}{\includegraphics*[angle=-90]{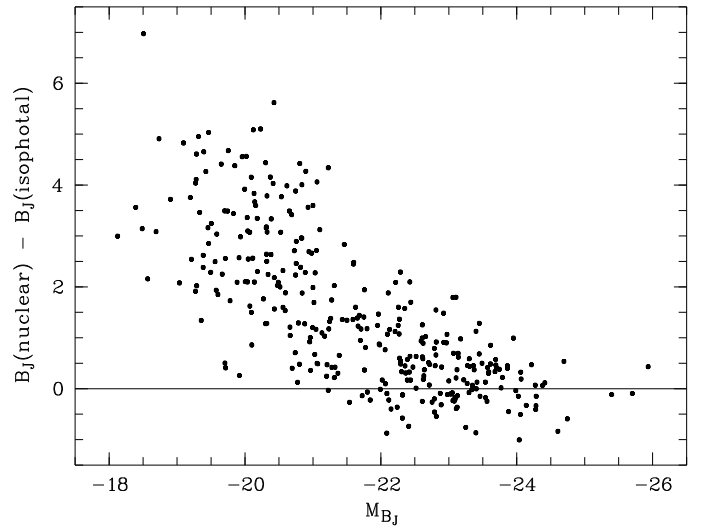}}
\caption{Difference between `isophotal DSS' and `nuclear' magnitudes, plotted against absolute nuclear magnitude $M_{B_J}$.}
\label{fig:mag_comp}
\end{figure}

\begin{figure*}
\includegraphics*[height=16cm,angle=-90]{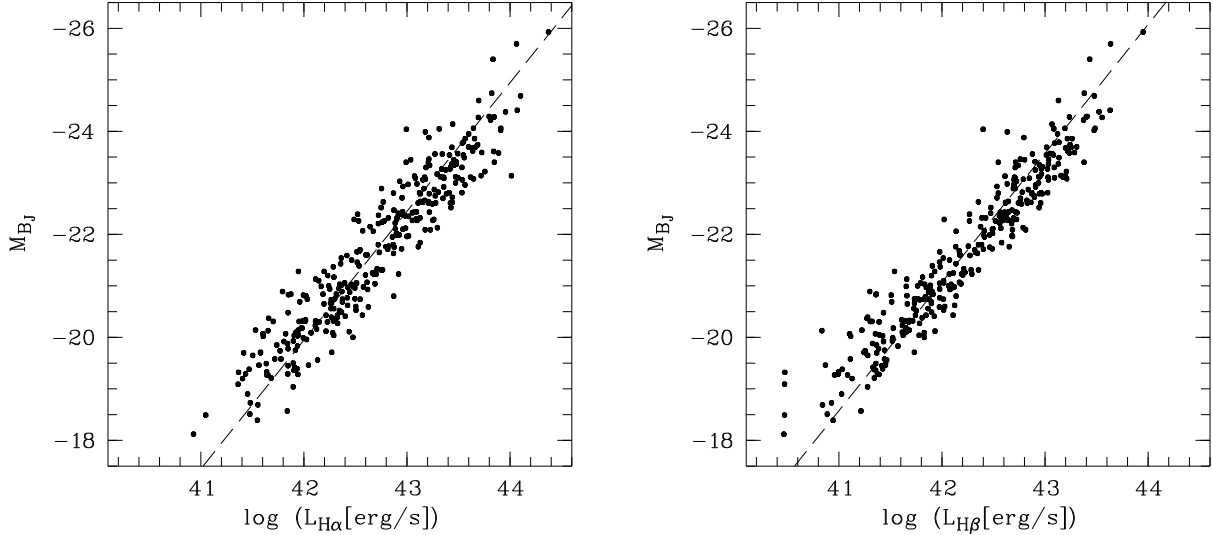}
\caption{Relation between absolute magnitudes in the $B_J$ band and
line luminosities (left: H$\alpha$, right: H$\beta$). A fixed-slope relation $M_{B_J}\propto -2.5\,\log L$ is shown as a dashed line in each panel, for illustration purposes only.}
\label{fig:luminosities}
\end{figure*}

\section{AGN luminosities        \label{sec:lummag}}

A long-standing issue for the determination of the local AGN luminosity function is the problem of how to disentangle nuclear AGN and host galaxy luminosities. Using total or isophotal photometric measurements will inevitably lead to luminosity- and redshift-dependent biases. For example,  the Seyfert luminosity functions determined in some earlier studies \citep[e.g.][]{Huchra:1992,Ulvestad:2001} clearly reflected more the distribution of host galaxy luminosities than AGN properties. Ideally, AGN and host light should be properly decomposed object by object; but as this would require high angular resolution data for all objects in a sample, such a route is presently not possible.

As a simplified approach to tackle this problem, we introduced in Sect.~\ref{sec:phot} our concept of `nuclear magnitudes' measured in the HES spectral plates. We did not subtract any host galaxy contribution, but we kept it to a minimum by measuring only the flux contribution of a nuclear point source. In Fig. ~\ref{fig:mag_comp} we compare these `nuclear' magnitudes with the more standard isophotal measurements in the DSS direct images. While for high-luminosity quasars (and for all quasars at higher redshifts, not shown in the figure), these two magnitudes give completely consistent results, there is an obvious discrepancy which increases towards lower luminosities. Clearly, the isophotal magnitudes are biased for almost all AGN with nuclear magnitudes fainter than $-23$, and useless for low-luminosity Seyfert galaxies. 

Hence it appears that the HES magnitudes are better estimates of nuclear luminosities than standard isophotal ones; but are they good enough? To investigate this we now compare the HES magnitudes with the luminosities of the broad emission lines. As H$\alpha$ and H$\beta$ are pure recombination lines, their luminosities should be proportional to the UV continuum \citep[e.g.][]{Yee:1980}. In an AGN spectrum, the fluxes of the broad lines are usually conspicuous and can be reasonably well measured even when the host galaxy contribution is strong or dominant. We thus adopt the Balmer emission line luminosities as proxies for the UV continuum luminosity of the AGN, without any host contribution.

In Fig.~\ref{fig:luminosities} we compare the HES broad band nuclear $B_J$ magnitudes with the luminosities in both Balmer lines.  The correlation is excellent and, more importantly, it extends over the entire range of luminosities.  To guide the eye, the dashed lines in Fig.~\ref{fig:luminosities} represent fixed-slope relations $M_{B_J}\propto -2.5\,\log L$. We see that the data come very close to a slope of unity in the case of $M_{B_J}$ vs.\ H$\alpha$, while the relation is slightly shallower for H$\beta$.

If the HES magnitudes were systematically affected by host galaxy contributions we should expect to see a saturation of $M_{B_J}$ values at small $L$. In fact no clear such trend is visible in the data, except maybe a small excess of a few points above the linear relation in the lower left corner of the right-hand panel. If at all, these few objects appear to be the only ones significantly affected by host galaxy light when described by the HES magnitudes. Overall we conclude that the broad band photometry of the HES is, in good approximation, a measure of the pure AGN luminosities.

\section{Luminosity functions}

\subsection{Luminosity function parameterisation}

The AGN luminosity function (LF) $\phi(L)$ is defined as the number of AGN per unit volume, per unit luminosity. The number of AGN per unit volume and per unit logarithmic luminosity $\Phi(L)=\mathrm{d}\Psi/\mathrm{d}(\log_{10}L)$ is given by $\Phi(L)=(L/\log_{10}e)\phi(L)$, where $\Psi(L)$ is the cumulative luminosity function. 

In the following we present the results in two different ways. We first estimate the LF in discrete luminosity bins, expressing it in the logarithmic form $\Phi(L)$ (or in the equivalent form in magnitudes). We then show the results of fitting these binned LFs with simple parametric expressions. As usual, the fit parameters are always expressed in terms of the non-logarithmic form $\phi(L)$. 

The most frequently adopted parametric form for the AGN luminosity function is a double power law:
\begin{equation}
\phi(L) = \frac{\phi^*/L_*}{(L / L_*)^{-\alpha} + (L / L_*)^{-\beta} } \ ,
\end{equation}
where $L_*$ is a characteristic break luminosity, $\phi^*$ the normalisation and $\alpha$ and $\beta$ are the two slopes. 

It will be seen that the local AGN LF is close to a single power law, so we will also consider that even simpler form:
\begin{equation}
\phi(L) = \frac{\phi^*}{L_*}\left( \frac{L}{L_*}\right) ^{\alpha} \ .
\end{equation}

Expressed in absolute magnitudes these functions have following form:
\begin{equation}
\Phi(M) = \frac{\Phi^*}{10^{0.4(1+\alpha)(M - M_*)} + 10^{0.4(1+\beta)(M - M_*)}}
\end{equation}
for the double power law, and
\begin{equation}
\Phi(M) = \Phi^*10^{-0.4(1+\alpha)(M - M_*)}
\end{equation}
for the single power law, with $\Phi^*=0.4\,\phi^*\ln(10)$ for both functions.

\begin{figure}
\resizebox{\hsize}{!}{\includegraphics*[height=18cm,angle=-90]{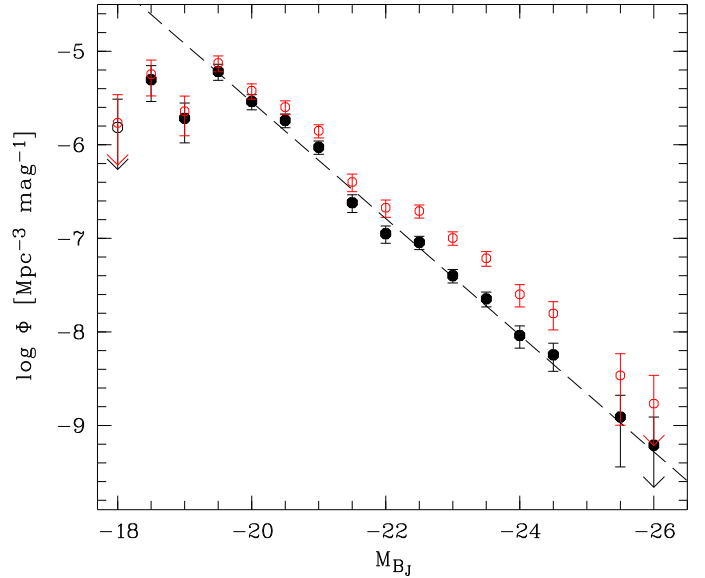}}
\caption{The differential broad band quasar luminosity function of the HES for $0.01 \leq z \leq 0.3$. The error bars are based on Poisson statistics. The arrow indicates a bin with only a single object. The black filled symbols show the luminosity function corrected for evolution using a simple PDE model;  red open symbols show the LF not corrected for evolution. The dashed line shows a single power law fit to the data.
}
\label{fig:qlfM}
\end{figure}

\subsection{Broad band Luminosity Function     \label{sec:BBLF}}

We first present the broad band luminosity function, using the full sample of 329 type-1 AGN with $z<0.3$. This updates the previous work of \citet{Wisotzki:2000a}, with several improvements: (i) The sample size has doubled. (ii) The quality of the external photometric calibration is improved. (iii)  The correction for Galactic extinction has been updated. (iv) We now also include the effects of differential evolution within the redshift range $0 < z < 0.3$ (see below).

We determined the binned luminosity function using the classical $V/V_\mathrm{max}$ estimator \citep{Schmidt:1968}. The luminosity function is then calculated by:
\begin{equation}
	\Phi(M) 
	= \frac{1}{\Delta M}\sum_{M - \Delta M / 2}^{M + \Delta M /2} \frac{1}{V_{\mathrm{max}}^k} 
 \label{eq:QLF} \ ,
\end{equation} 
where $\Delta M$ is the bin size and $V_{\mathrm{max}}$ is the survey volume in which the object $k$ could have been detected within the flux limit of the survey and the given redshift range. Recall that the dispersion in limiting magnitudes over the 380 HES fields is taken into account by the magnitude dependence of the effective survey area $\Omega_{\mathrm{eff}}$ (Fig.~\ref{fig:effA}). Thus $V_{\mathrm{max}}$ is given by
\begin{equation} 
	V_\mathrm{max} 
	= \int_{z_\mathrm{min}}^{z_\mathrm{max}} \Omega_{\mathrm{eff}}(m) \frac{dV}{dz} dz  
	\label{eq:vmax} \ .
\end{equation}
In adopting this form we explicitly assume that the probability of finding an AGN in the HES is independent of redshift. While this would be too strong an assumption for the full quasar sample, it is certainly justified for the restricted low-redshift range $z<0.3$ considered here. The SEDs of typical type~1 AGN (without host galaxy contributions) at such low redshifts are distinctly blue in the optical/UV, and significant marked variation with redshift occur beyond $z>0.5$. (Note that we do not consider here the role of `red' quasars which would be lost in the HES altogether.) If $\mathcal{S}(m,z)$ is the redshift- and magnitude-dependent survey selection function, we can safely marginalise over redshifts and set $\mathcal{S}(m,z) \propto \Omega_{\mathrm{eff}}(m)$. 
We reiterate that the effective area $\Omega_{\mathrm{eff}}(m)$ already fully accounts for the `target sampling' incompleteness due to photometric losses, overlapping spectra etc.\ which has been carefully determined in the same way as described by \citet{Wisotzki:2000c}. Furthermore, as essentially 100\% of the follow-up spectra of the HES could be classified with high fidelity, we assume that the probability of selecting an AGN within the targeted redshift range and brighter than the flux limit is unity.

The resulting AGN luminosity function is shown in Fig.~\ref{fig:qlfM}, covering the range $-26 \leq M_{B_J} \leq -18$ in bins of 0.5~mag. It shows remarkably little structure and rises steadily up to the faintest luminosities in the sample. At $M_{B_J} \ga -19$ there
appears to be an abrupt break which almost certainly is an artefact, indicating the inevitable onset of severe incompleteness in the HES sample for very low luminosities. For such objects, the host galaxy contribution even to the HES nuclear extraction scheme will be substantial, modifying the slitless spectra in a way that they no longer can be discriminated from normal, inactive galaxies. It is remarkable that this effect plays a role only for the lowest luminosity bins;  there is no gradual turnover that might indicate incompleteness already at higher luminosities (alternatively, invoking incompleteness would imply an even steeper LF which would be inconsistent with other results, see below).

If we ignore the lowest luminosity bins affected by incompleteness, the $M_{B_J}$ AGN luminosity function is consistently described by a single power law with slope $\alpha=-2.4$. Fitting instead a double power law to all bins results in the same bright-end slope and a break at  $M_{B_J}=-18.75$. A double power law fit to these data is apparently not physically meaningful. 

The observation that the local AGNLF is perfectly described by a single power law is in excellent qualitative and quantitative agreement with previous results obtained in the course of the HES \citep{Koehler:1997,Wisotzki:2000a}. It is however inconsistent with $z=0$ extrapolations of the double power law AGNLF obtained at higher redshifts. We will discuss this point in section~\ref{sec:down}.

\begin{table}
\caption{Binned differential luminosity function in the $B_J$ band, corrected for evolution. $N$ is the number of objects contributing per bin. The luminosity function is expressed as number density per Mpc$^3$ per unit magnitude.}
\label{tab:qlfM}
\centering
\begin{tabular}{rrcll} \hline\hline \noalign{\smallskip}
$M_{B_J}$ & $N$ & $\log \Phi(M_{B_J})$ & \multicolumn{2}{c}{$\sigma(\log \Phi)$}\\ 
\noalign{\smallskip} \hline \noalign{\smallskip}
$-$18.0 & 1 & $-$5.81 & +0.3  & $-\infty$  \\ 
$-$18.5 & 6 & $-$5.30 & +0.15 & $-$0.24  \\ 
$-$19.0 & 5 & $-$5.72 & +0.17 & $-$0.26  \\ 
$-$19.5 & 28 & $-$5.22 & +0.08 & $-$0.09  \\ 
$-$20.0 & 30 & $-$5.54 & +0.08 & $-$0.09  \\ 
$-$20.5 & 37 & $-$5.74 & +0.07 & $-$0.08  \\ 
$-$21.0 & 40 & $-$6.03 & +0.07 & $-$0.07  \\ 
$-$21.5 & 23 & $-$6.62 & +0.09 & $-$0.10  \\ 
$-$22.0 & 24 & $-$6.95 & +0.08 & $-$0.10  \\ 
$-$22.5 & 40 & $-$7.04 & +0.06 & $-$0.08  \\ 
$-$23.0 & 38 & $-$7.40 & +0.07 & $-$0.08  \\ 
$-$23.5 & 31 & $-$7.65 & +0.08 & $-$0.08  \\ 
$-$24.0 & 14 & $-$8.04 & +0.10 & $-$0.13  \\ 
$-$24.5 & 9 & $-$8.25 & +0.13 & $-$0.17  \\ 
$-$25.5 & 2 & $-$8.91 & +0.23 & $-$0.53  \\ 
$-$26.0 & 1 & $-$9.21 & +0.3  & $-\infty$  \\ 
\noalign{\smallskip} \hline
\end{tabular} 
\end{table}

The evolution of comoving AGN space densities with redshift is sufficiently fast that there is a noticeable effect even within the range $0<z<0.3$.  To derive a truly `local' ($z=0$) AGNLF we have to take evolution into account. The $V/V_\mathrm{max}$ formalism \citep{Schmidt:1968} provides a simple but adequate recipe to do so.  If our sample were unaffected by evolution, we would expect to find $\langle V/V_\mathrm{max}\rangle=0.5$. Our measured value is $\langle V/V_\mathrm{max}\rangle = 0.54 \pm 0.02$, implying some evolution. To correct the $z<0.3$ AGNLF to $z=0$, we approximate the evolution within this small redshift interval as pure density evolution (PDE), i.e. $\Phi(z) \propto (1+z)^{k_D}$. We varied the density evolution parameter $k_D$ until we reached $\langle V/V_\mathrm{max}\rangle \approx0.5$. To increase the leverage we performed the same exercise for a larger redshift range, including quasars from the HES sample up to $z=0.6$. We found that the evolution within this redshift interval is well described by a PDE model with $k_D=5$, essentially independent of the exact value of the outer redshift boundary. Notice that beyond the very local universe, the HES samples only the brightest quasars, and our correction for the most part concerns these high-luminosity bins only. For a single power law, density and luminosity evolution are indistinguishable. Therefore our evolution correction does not critically depend on the choice of the actually adopted model.

We then applied the parameterised density evolution to the HES $z<0.3$ sample to recompute the evolution-corrected $z=0$ AGNLF. Note that we used the objects at $z>0.3$ only to constrain the density evolution index $k_D$ and not for the luminosity function.

We still obtain a relation very close to a single power law, which however is slightly steeper than the uncorrected one. The best fit power law slope is now $\alpha = -2.6$. This evolution-corrected AGNLF is also shown in Fig.~\ref{fig:qlfM}. It is provided in tabulated form in Table~\ref{tab:qlfM}.

\subsection{Emission Line Luminosity Function       \label{sec:ELLF}}

The emission line luminosity function (ELF) -- the number of AGN per unit volume per unit logarithmic emission line luminosity -- is given by
\begin{equation}
\Phi(L) = \frac{1}{\Delta \log L}\sum_{k} \frac{1}{V_{\mathrm{max}}^k}  \ .
\end{equation}
The selection of AGN in the HES in this redshift range (and up to $z \sim 2.5$) is exclusively based on continuum SED properties and independent of emission line properties \citep{Wisotzki:2000c}. Thus the selection function is the same as for the broad band luminosity function, and the $V_{\mathrm{max}}$ values needed for the determination of the emission line luminosity function are also the same as for the $M_{B_J}$ luminosity function, except for a correction factor containing the  spectroscopic incompleteness. As discussed in \S~\ref{sec:sample}, our sample has an overall spectroscopic completeness of 324/329 $\approx$ 98.5~\%.  Of the 324 objects with spectra, 6 do not show H$\beta$ and 29 do not show H$\alpha$, due to insufficient spectroscopic coverage or atmospheric absorption. This incompleteness does not affect any particular object types preferentially, and we assumed the losses to be randomly distributed. We therefore adopted the effective survey area as before, multiplied by a factor of 318/329 for H$\beta$ and by 295/329 for H$\alpha$, respectively. Unless explicitly stating otherwise, we always refer to the evolution-corrected $V_{\mathrm{max}}$ values, thus providing luminosity functions valid for exactly $z=0$.

\begin{figure}
\resizebox{\hsize}{!}{\includegraphics*[angle=-90]{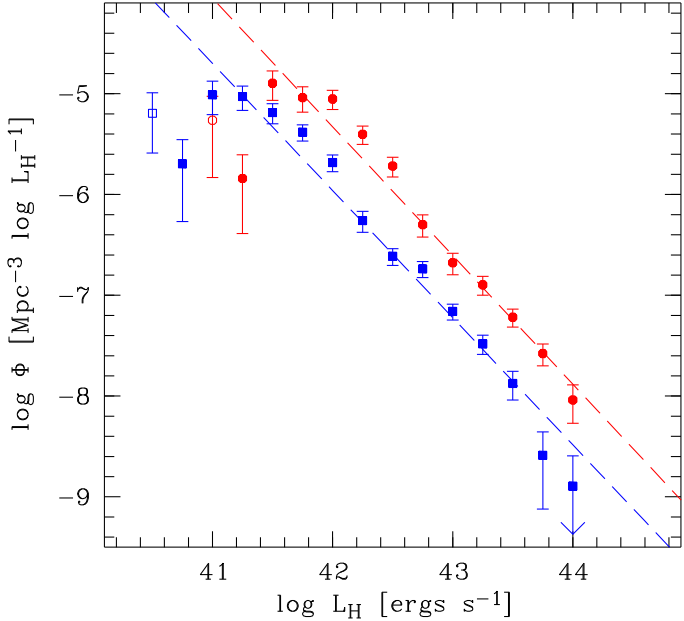}}
\caption{Binned emission line AGN luminosity functions, for H$\alpha$ (red filled circles) and H$\beta$ (blue squares). Open symbols indicate incompletely filled bins. The dashed lines show the best fit single power law to the H$\alpha$ and H$\beta$ data, respectively.}
\label{fig:qlfL}
\end{figure}

The resulting ELFs for H$\alpha$ and H$\beta$ are shown in Fig.~\ref{fig:qlfL}, binned into luminosity intervals of 0.25~dex. The binned values are given in Table~\ref{tab:qlfL}. 

\begin{table}
\caption{Binned H$\alpha$ and H$\beta$ emission line luminosity functions.}
%\normalsize
\label{tab:qlfL}
\centering
\begin{tabular}{lrrrr} \hline \hline \noalign{\smallskip} \noalign{\smallskip} 
  & \multicolumn{2}{c}{\normalsize{\textbf{H}$\mathbf{\alpha}$}}	& 
\multicolumn{2}{c}{\normalsize{\textbf{H}$\mathbf{\beta }$}} \\[0.5ex]
$\log L$ & $N$ & $\log \Phi(\mathrm{H}\alpha )$  & $N$ & $\log \Phi(\mathrm{H}\beta)$ \\ 
\noalign{\smallskip} \hline \noalign{\smallskip}
40.5 &  &  & 4 & $-$5.19$^{+0.2}_{-0.4}$ \\  \noalign{\smallskip}
40.75 &  &  & 3 & $-$5.7$^{+0.24}_{-0.57}$ \\ \noalign{\smallskip}
41.0 & 2 & $-$5.26$^{+0.23}_{-0.57}$ & 11 & $-$5.01$^{+0.13}_{-0.2}$ \\ \noalign{\smallskip}
41.25 & 2 & $-$5.84$^{+0.23}_{-0.55}$ & 23 & $-$5.03$^{+0.11}_{-0.14}$ \\ \noalign{\smallskip}
41.5 & 16 & $-$4.9$^{+0.13}_{-0.17}$ & 26 & $-$5.19$^{+0.09}_{-0.11}$ \\ \noalign{\smallskip}
41.75 & 23 & $-$5.04$^{+0.11}_{-0.14}$ & 38 & $-$5.38$^{+0.07}_{-0.09}$ \\ \noalign{\smallskip}
42.0 & 34 & $-$5.05$^{+0.08}_{-0.11}$ & 37 & $-$5.68$^{+0.07}_{-0.09}$ \\ \noalign{\smallskip}
42.25 & 38 & $-$5.4$^{+0.08}_{-0.1}$ & 25 & $-$6.26$^{+0.09}_{-0.12}$ \\ \noalign{\smallskip}
42.5 & 35 & $-$5.72$^{+0.09}_{-0.11}$ & 36 & $-$6.61$^{+0.07}_{-0.09}$ \\ \noalign{\smallskip}
42.75 & 28 & $-$6.3$^{+0.1}_{-0.12}$ & 45 & $-$6.74$^{+0.07}_{-0.08}$ \\ \noalign{\smallskip}
43.0 & 32 & $-$6.68$^{+0.1}_{-0.12}$ & 35 & $-$7.16$^{+0.07}_{-0.09}$ \\ \noalign{\smallskip}
43.25 & 34 & $-$6.9$^{+0.09}_{-0.1}$ & 22 & $-$7.48$^{+0.08}_{-0.11}$ \\ \noalign{\smallskip}
43.5 & 28 & $-$7.22$^{+0.08}_{-0.1}$ & 10 & $-$7.87$^{+0.11}_{-0.17}$ \\ \noalign{\smallskip}
43.75 & 17 & $-$7.58$^{+0.1}_{-0.12}$ & 2 & $-$8.59$^{+0.23}_{-0.53}$ \\ \noalign{\smallskip}
44.0 & 6 & $-$8.04$^{+0.15}_{-0.23}$ & 1 & $-$8.89$^{+0.3}_{-\infty}$ \\ 
\noalign{\smallskip} \hline
\end{tabular} 
\end{table}

\begin{table*}
\caption{Fit parameters for the AGNLF.}
\label{tab:qlf}
\centering
\begin{tabular}{rrrrrrrr}
\hline \hline \noalign{\smallskip}
AGNLF & & $\phi^\ast$, $\Phi^\ast$ [Mpc]$^{-3}$ & $M_\ast$, $\log L_\ast$ & $\beta$ & $\alpha$  & $\chi^2$ & $\chi^2$/dof \\ 
\noalign{\smallskip} \hline \noalign{\smallskip}
$M_{B_J}$ & $z<0.3$ & $1.80\times 10^{-7}$ & $-$22.44 & $-$2.35 & -- & 16.08  & 1.61 \\ 
$M_{B_J}$ & $z=0$ & $8.36\times 10^{-8}$ & $-$22.46 & $-$2.56 & -- & 14.56  & 1.46 \\ 
H$\alpha$ & $z=0$ & $2.18\times 10^{-7}$ & 42.76 & $-$2.28 & -- & 18.51  & 2.31 \\ 
H$\beta$ & $z=0$ & $1.07\times 10^{-7}$ & 42.51 & $-$2.26 & -- & 34.17  & 3.42 \\ 
H$\alpha$ + SDSS & $z=0$ & $5.59\times 10^{-6}$ & 41.58 & -2.21 & -- & 97.0  & 2.85 \\ 
H$\alpha$ + SDSS & $z=0$ & $1.8\times 10^{-5}$ & 41.67 & -2.58 & $-$1.80  & 37.32  & 1.49 \\ 
\noalign{\smallskip} \hline \noalign{\smallskip}
$M_{B}$ (HES+SDSS) & $z=0$ & $5.60\times 10^{-6}$ & $-$19.25 & $-$2.40 & $-$ & 65.5  &  4.09 \\ 
$M_{B}$ (HES+SDSS) & $z=0$ & $1.55\times 10^{-5}$ & $-$19.46 & $-$2.82 & $-$2.00 & 24.4  &  1.62 \\ 
$\log L_\mathrm{Bol}$ (HES+SDSS) & $z=0$ & $2.22\times 10^{-5}$ & 44.55 & $-$2.92 & $-$2.17 & 25.8  &  1.72 \\
\noalign{\smallskip} \hline
\end{tabular}
\end{table*}

Figure~\ref{fig:qlfL} displays a very similar behaviour of the ELF when compared to the broad-band LF of Fig.~\ref{fig:qlfM}: It rises nearly as a straight line, i.e.\ as a single power law, until a sharp cutoff at low luminosities indicates the onset of sample selection incompleteness. Fitting power law relations to the data (again excluding the obviously incomplete lowest luminosity bins) gives slopes of $\alpha_\mathrm{H\alpha} = -2.28$ and $\alpha_\mathrm{H\beta} = -2.26$, respectively. Fitting a double power law improves the fit quality only marginally. We conclude that a description of the ELF as a single power law, for the luminosity range covered by our data, seems most appropriate.

\section{Discussion} \label{sec:compL}

\subsection{Low luminosity-AGN: Comparison with SDSS   \label{sec:sdss}}

The sharp cutoff in the binned luminosity functions at nuclear luminosities $M_{B_J} \ga -19$ or $\log L_{H\alpha} \la 42$ clearly signals the onset of incompleteness in our sample. This luminosity approximately marks the limit where AGN cease to be conspicuous in the optical and tend to be masked by their host galaxies. However, deep spectroscopic surveys of galaxies have shown  that the AGN phenomenon persists down to very low levels \citep[e.g.][]{Ho:1997}. In those cases, the only traceable indicator of nuclear activity in the optical are the emission lines, thus the statistics have to be expressed in terms of an ELF. This was recently performed by \citet[][hereafter H05]{Hao:2005}, who selected a set of $\sim$\,1000 Seyfert~1 galaxies from the Sloan Digital Sky Survey (main galaxy sample) to measure H$\alpha$ line luminosities and construct the ELF. The redshift range covered in their sample is $0<z<0.15$, and the luminosity range is ($10^{38.5}$--$10^{43}$) erg~s$^{-1}$. H05 found their ELF to be in good agreement with several parametric descriptions, including single and double power laws and also a Schechter function. However, differences between these forms become manifest only at their highest luminosities, for $L(\mathrm{H}\alpha)\sim 10^{42}$~erg~s$^{-1}$ . Over much of the luminosity range, their data suggest a single power law.

\begin{figure}
\resizebox{\hsize}{!}{\includegraphics*[angle=-90]{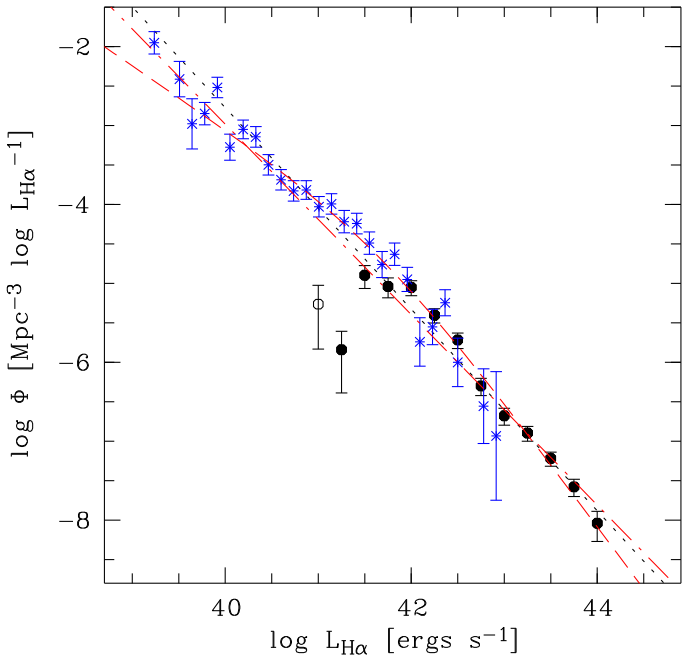}}
\caption{Comparison of the H$\alpha$ luminosity function derived in this work (black points), with the work of \citet{Hao:2005} (blue asterisks). The black dotted line gives the best fit single power law to our H$\alpha$ luminosity function. The red dashed dotted line gives the best fit power law to the combined data set, whereas the red dashed line gives the best fit double power law.}
\label{fig:compqlfL}
\end{figure}

Fortunately the \emph{high-luminosity end} of H05 overlaps quite well with the \emph{low-luminosity end} of our H$\alpha$ ELF, so that a comparison is straightforward.  This is shown in Fig.~\ref{fig:compqlfL} where we plot the Seyfert~1 H$\alpha$ luminosity function of H05 (adapted to our cosmology) together with our H$\alpha$ LF. The transition from one dataset to the other is remarkably smooth, if the incomplete lowest luminosity bins in the HES sample are ignored. Over the luminosity range in common, both ELFs are fully  consistent with each other. 

Even more remarkably, the shape of the combined H$\alpha$ LF -- now covering almost 5 orders of magnitude in luminosity -- is still close to that of a single power law. Already the fit to the HES H$\alpha$ LF alone is almost consistent with the H05 LF. Combining both samples, we find a best-fit power law slope of  $\alpha \approx -2.2$. This fit is shown as the dashed-dotted line in Fig.~\ref{fig:compqlfL}. 

However, there is evidence for \emph{some} curvature in the LF, as the H05 LF alone is flatter ($\alpha = -2.02$) than the HES ELF. This is manifest in a `bulge' around $10^{42}$ erg s$^{-1}$, where the space density of the combined LF is above the fitted single power law. A better fit is obtained by a double power law breaking at $\log L_\star=41.67$, with a faint-end slope of $\alpha=-1.8$ describing the SDSS data and a bright-end slope of $\beta = -2.6$ describing the HES. This mildly curved relation, shown by the dashed line in Fig.~\ref{fig:compqlfL}, traces the combined data extremely well.

Some caveats are in place regarding the combination of the two datasets. Although H05 determined the narrow H$\alpha$ components separately, their broad-line ELF does not have this component subtracted, whereas we tried to remove it. As such a removal was possible in only a small number of cases, the different treatment does not make much difference. More relevant might be the possibility of a systematic variation of the narrow H$\alpha$ contribution to the total H$\alpha$ flux with luminosity. However, the \emph{narrow} H$\alpha$ LF  published by H05 has also a slope of $-1.8$, which indicates that there should be no major bias introduced.

Another methodical difference lies in the fact that we corrected our LF for evolution, whereas H05 did not. If we assume that the most luminous objects of H05 lie close to their high-redshift limit of $z< 0.15$ and adopting our simple PDE recipe, then the H05 space densities would have to be corrected downward by a factor of $1.15^5 \approx 2$, i.e.\ by 0.3~dex; this correction would rapidly decrease towards lower luminosities. The net effect would hardly be visible in Fig.~\ref{fig:compqlfL}.

We note that recently, \citet{Greene:2007a} (hereafter GH07) derived the AGN H$\alpha$ LF based on a combined sample of about 9000 broad line AGN from the SDSS. Their space densities are considerably below ours, and the two LFs are highly inconsistent (the same discrepancy exists between the GH07 and H05 results). This inconsistency has been traced back to an error made by GH07 in the determination of their $V/V_{\mathrm{max}}$ values (J.~Greene, private communication). Therefore the luminosity function as well as the black hole mass function presented in GH07 are incorrect. Removing the error alleviates the discrepancy, and the corrected H$\alpha$ luminosity function for the low redshift AGN sample from GH07 is consistent with the HES luminosity function presented in this work.

\begin{figure*}
\centering
\includegraphics*[height=14cm,angle=-90]{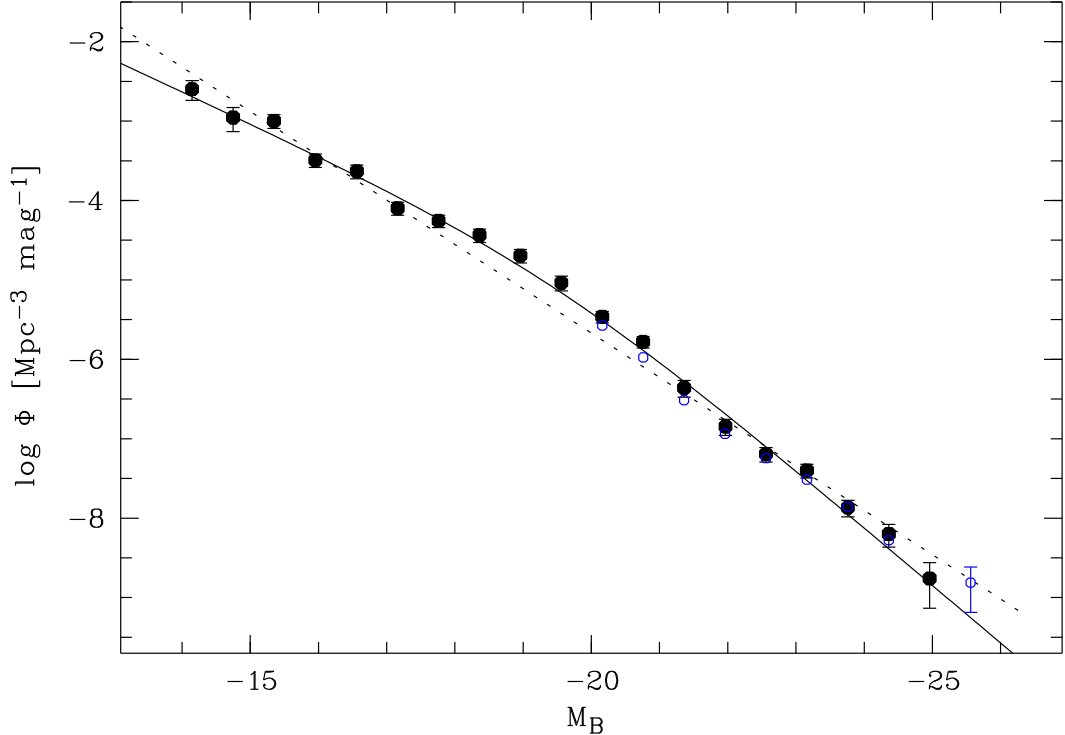}
\caption{Local AGN luminosity function constructed as a combination of the SDSS broad line AGN LF from \citet{Hao:2005} and our  H$\alpha$ HES bright AGN LF, both converted to the $B$ band. The black solid line and dotted line show the double and single power law fits, respectively. The open blue circles show the directly determined broad band ($M_{B_J}$) LF for comparison.}
\label{fig:combinedLF}
\end{figure*}

\subsection{The combined local AGN luminosity function} \label{sec:combinedLF}

Given the good agreement of the local HES LF, which traces the bright end, and the SDSS Seyfert~1 LF sampling the faint end, and having the mentioned caveats in mind, it is justified to combine both datasets. Our aim is to present a single best-knowledge local AGNLF in the optical, covering the broadest possible range of luminosities. Because of its robustness against dilution due to host galaxy light, we used the H$\alpha$ luminosity functions of HES and SDSS. We adopted a bin size of 0.24~dex (0.6~mag) as an integer multiple of the binned values published by H05. We then recomputed the H$\alpha$ ELF from the HES data for the same bins and merged the two datasets, with weights provided by the inverse statistical variances. 

For easy comparison with other AGN luminosity functions, especially at higher redshifts, we converted this combined LF into two common reference systems: (i) absolute magnitudes $M_B$ in the standard Johnson $B$ band; and (ii) bolometric luminosity units.
These conversions involve translating the H$\alpha$ data into broad band or bolometric fluxes. There is a very tight correlation between H$\alpha$ luminosity and absolute blue magnitude in the HES (Fig.~\ref{fig:luminosities}). The translation relation is $M_B= -2.1\,(\log L(\mathrm{H}\alpha)-42)-20.1$. This relation is covered by our data down to $M_{B} \simeq -19$. In order to incorporate also the lower luminosity SDSS data we now make the somewhat unguarded step of extrapolating the translation of $L_\mathrm{H\alpha}$ to $M_B$ towards lower $L$. This certainly introduces additional uncertainties, including the possibility that some of the lowest luminosity AGN could have very different spectral energy distributions (for example, the structure of the accretion disk might change drastically). On the other hand, there is no reason to expect such a change to occur just at the transition luminosity from HES to SDSS, so some degree of extrapolation is most probably justified. 

The resulting combined local AGN LF is shown in Fig.~\ref{fig:combinedLF}, ranging from $M_B \ga -25$ to $M_B \la -15$ (with the above caveat). Also shown as a dotted line is the best fit single power law, and as a solid line the best fit double power law; the fit parameters are provided in Table~\ref{tab:qlf}. The double power law gives a very good overall description, but the departure from a single power law is not large, albeit statistically significant. There are some minor wiggles in the binned LF that are most probably due to underlying unaccounted for systematics; note however that there is no trace of the HES-SDSS intersection (rather: transition region) around $M_B \sim -20$.

For the convenience of the reader we also provide this LF in tabulated form, both in terms of the $B$ band and as a bolometric LF. For the latter we adopted the luminosity-dependent bolometric corrections of \citet{Marconi:2004}. A separate double power law fit to the resulting bolometric LF is also provided in Table~\ref{tab:qlf}. Recall that this bolometric luminosity function is valid only for broad-line (type~1) AGN, without any accounting for obscuration.

\begin{table}
\caption{Combined binned local AGNLF, based on the SDSS broad line galaxy sample (faint end) and the Hamburg/ESO Survey (bright end). The bolometric AGNLF has been computed using the bolometric corrections by \citet{Marconi:2004}.}
%\normalsize
\label{tab:combLF}
\centering
\begin{tabular}{rrrr} \hline \hline \noalign{\smallskip} \noalign{\smallskip} 
$M_B$  & $\log \Phi(M_B)$  & $\log L_\mathrm{Bol}$ & $\log \Phi(L_\mathrm{Bol})$ \\ 
\noalign{\smallskip} \hline \noalign{\smallskip}
$-$14.2 & $-$2.60$^{+0.11}_{-0.14}$ & 42.63 & $-$2.12$^{+0.10}_{-0.15}$ \\ \noalign{\smallskip} 
$-$14.8 & $-$2.95$^{+0.12}_{-0.18}$ & 42.83 & $-$2.48$^{+0.13}_{-0.18}$ \\ \noalign{\smallskip} 
$-$15.4 & $-$3.00$^{+0.08}_{-0.09}$ & 43.03 & $-$2.52$^{+0.07}_{-0.1}$ \\ \noalign{\smallskip} 
$-$16.0 & $-$3.49$^{+0.07}_{-0.09}$ & 43.24 & $-$3.02$^{+0.07}_{-0.09}$ \\ \noalign{\smallskip} 
$-$16.6 & $-$3.63$^{+0.08}_{-0.10}$ & 43.44 & $-$3.16$^{+0.07}_{-0.10}$ \\ \noalign{\smallskip} 
$-$17.2 & $-$4.10$^{+0.08}_{-0.09}$ & 43.65 & $-$3.64$^{+0.08}_{-0.08}$ \\ \noalign{\smallskip} 
$-$17.8 & $-$4.26$^{+0.08}_{-0.08}$ & 43.86 & $-$3.80$^{+0.07}_{-0.08}$ \\ \noalign{\smallskip} 
$-$18.4 & $-$4.44$^{+0.08}_{-0.09}$ & 44.07 & $-$3.98$^{+0.07}_{-0.10}$ \\ \noalign{\smallskip} 
$-$19.0 & $-$4.69$^{+0.07}_{-0.10}$ & 44.28 & $-$4.24$^{+0.07}_{-0.10}$ \\ \noalign{\smallskip} 
$-$19.6 & $-$5.04$^{+0.09}_{-0.10}$ & 44.49 & $-$4.59$^{+0.08}_{-0.10}$ \\ \noalign{\smallskip} 
$-$20.2 & $-$5.46$^{+0.06}_{-0.08}$ & 44.71 & $-$5.02$^{+0.06}_{-0.08}$ \\ \noalign{\smallskip} 
$-$20.8 & $-$5.78$^{+0.07}_{-0.08}$ & 44.93 & $-$5.34$^{+0.06}_{-0.08}$ \\ \noalign{\smallskip} 
$-$21.4 & $-$6.36$^{+0.09}_{-0.12}$ & 45.15 & $-$5.93$^{+0.10}_{-0.11}$ \\ \noalign{\smallskip} 
$-$22.0 & $-$6.85$^{+0.09}_{-0.11}$ & 45.38 & $-$6.42$^{+0.09}_{-0.11}$ \\ \noalign{\smallskip} 
$-$22.6 & $-$7.19$^{+0.08}_{-0.10}$ & 45.60 & $-$6.77$^{+0.09}_{-0.10}$ \\ \noalign{\smallskip} 
$-$23.2 & $-$7.40$^{+0.08}_{-0.09}$ & 45.82 & $-$6.98$^{+0.08}_{-0.09}$ \\ \noalign{\smallskip} 
$-$23.8 & $-$7.87$^{+0.10}_{-0.11}$ & 46.05 & $-$7.45$^{+0.10}_{-0.11}$ \\ \noalign{\smallskip} 
$-$24.4 & $-$8.20$^{+0.12}_{-0.16}$ & 46.28 & $-$7.78$^{+0.12}_{-0.17}$ \\ \noalign{\smallskip} 
$-$25.0 & $-$8.76$^{+0.20}_{-0.37}$ & 46.51 & $-$8.34$^{+0.19}_{-0.38}$ \\ 
\noalign{\smallskip} \hline
\end{tabular} 
\end{table}

\subsection{Comparison with X-ray selected samples         \label{sec:compX}}

X-ray surveys have made a great impact on our understanding of the AGN population, chiefly through their ability to find low-luminosity AGN at all redshifts. However, because of the expensive spectroscopic follow-up, sample statistics are still moderate despite considerable efforts.

The only dedicated effort to estimate an optical local AGNLF from an X-ray selected sample was published by \citet{Londish:2000}. Their bright-end slope of $\beta=-2.1$
(without evolution correction) is similar to ours, but they obtained a shallow faint end slope of $\alpha=-1.1$, however with considerable error bars. Our new results, in particular in combination with SDSS, show clearly that such a flat slope is ruled out and that the local AGNLF continues to rise towards very faint luminosities.

\begin{figure}
\centering
\resizebox{\hsize}{!}{\includegraphics*[angle=-90]{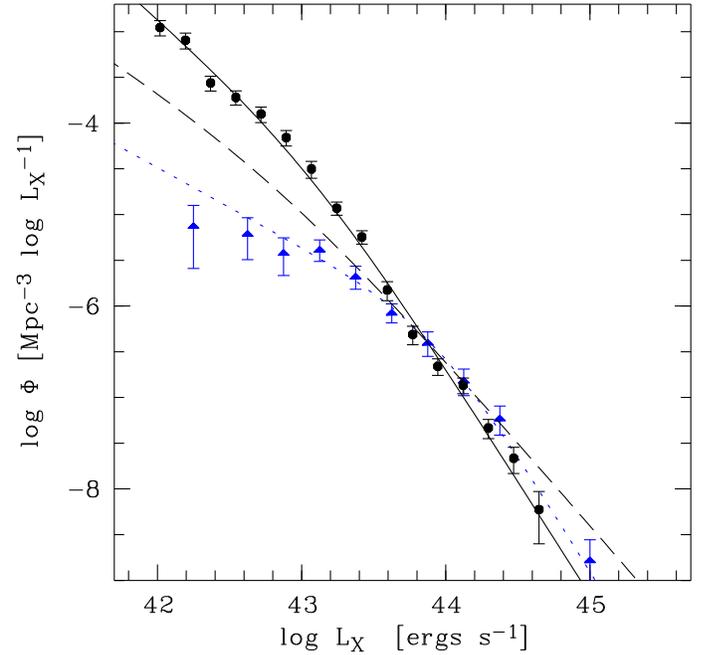}}
\caption{Comparison of the soft X-ray LF of type~1 AGN by \citet{Hasinger:2005} with the prediction based on the optical local AGNLF. The filled blue triangles show the binned X-ray (0.5--2~keV) LF within $z=0.015 - 0.2$, corrected to $z=0$. The blue dotted line shows the LDDE model of \citet{Hasinger:2005} for $z=0$. The filled black circles and the solid line show the binned data and the double power law fit to the optical HES+SDSS LF, converted to soft X-rays using a luminosity dependent correction. The dashed line shows the same double power law fit, but converted to soft X-rays using a constant correction.}
\label{fig:comp_xray}
\end{figure}

A local ($z=0$) luminosity function from an AGN sample selected in the soft X-ray (0.5--2~keV) band was presented by \citet{Hasinger:2005}. Their sample is restricted to unabsorbed type~1 AGN and is therefore very comparable to ours. To facilitate a comparison, we again used the $L$-dependent bolometric corrections of \citet{Marconi:2004} to convert our combined optical AGNLF into the soft X-ray domain. In Fig.~\ref{fig:comp_xray} we compare the $z=0$ XLF of \citet{Hasinger:2005} with our prediction. The X-ray points are represented by the filled blue triangles, which are the binned estimate in the redshift shell $z=0.015 - 0.2$ of Fig.~7 in \citet{Hasinger:2005}, corrected to redshift zero. The blue dotted line shows their best-fit luminosity-dependent density evolution (LDDE) model for $z=0$.

For intermediate and high luminosities, the observed XLF and the prediction based on the optical AGNLF are in good agreement. However, the two LFs disagree strongly at the faint end, with the optical LF predicting more than an order of magnitude higher space densities at given X-ray luminosity compared to the directly determined XLF.

Is there an explanation for these discrepancies? One possibility might be that the adopted conversion from optical to X-ray luminosities has been inadequate. In order to explore this option we alternatively tried a constant optical/X-ray luminosity ratio; the dashed line in Fig.~\ref{fig:comp_xray} shows our double-power law relation converted with such a relation. Evidently, that predicted LF is a very good match to the bright end of the XLF. At the faint end the two LFs still disagree, but the disagreement is now much less. Thus it appears possible that the luminosity dependence of the optical/X-ray ratio is much weaker than usually assumed; if so, it would certainly help to reconcile the two luminosity functions. Another possibility is, of course, incompleteness among the fainter objects in the X-ray sample. We reiterate however that since both optically and X-ray selected samples contain only broad-line AGN, any incompleteness due to obscuration should be irrelevant in this context.

\begin{figure}
\centering
\resizebox{\hsize}{!}{\includegraphics*[angle=-90]{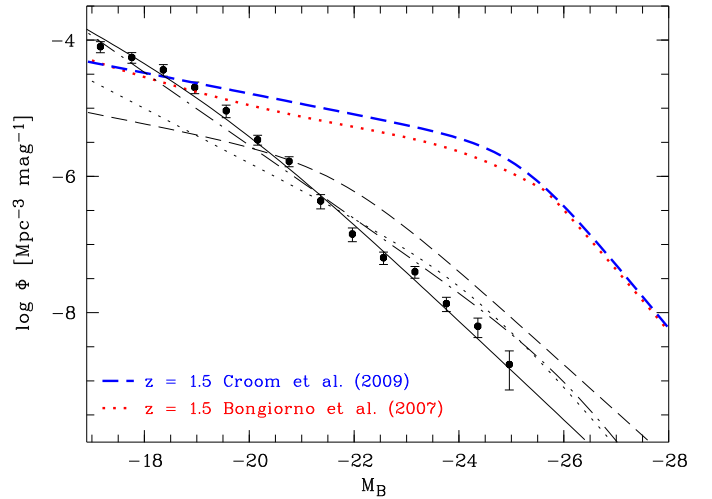}}
\caption{Shape evolution and `downsizing' of the AGNLF between $z=1.5$ and $z=0$. Filled circles and solid line denote our combined HES+SDSS AGNLF at $z=0$. The blue dashed and red dotted lines show the best fit model to the data of the $z=1.5$ AGNLF of the 2SLAQ+SDSS \citep{Croom:2009} and the VVDS+SDSS \citep{Bongiorno:2007}, respectively. The thin black dashed, dashed-dotted and dotted lines show the $z=0$ extrapolations of the LEDE model and the LDDE model by \citet{Croom:2009} and the LDDE model by \citet{Bongiorno:2007}, respectively. }
\label{fig:compqlf15}
\end{figure}

\subsection{Comparison with higher redshifts: Evidence for `AGN downsizing' \label{sec:down}}

We now perform a direct comparison of our local AGNLF with luminosity functions based on surveys that probe mainly the higher redshift AGN population. This is interesting not only because a comparison with $z=0$ provides the longest leverage in redshifts, but also because the local luminosity function can be traced to very faint luminosities and high space densities. A full investigation of the redshift evolution of the AGNLF is outside of the scope of this paper; we limit ourselves to a simple comparison between the $z=0$ LF determined above and parametric representations of the AGNLF evaluated at $z=1.5$, the latter chosen as a representative point at moderately high redshift where several surveys have been able to leave their marks. 

The results of this comparison are displayed in Fig.~\ref{fig:compqlf15}. The datapoints and the solid line show our combined local AGNLF. The thick dashed line shows the $z=1.5$ fit to the optical AGNLF, very recently obtained by \citet{Croom:2009}, combining QSOs from the 2SLAQ survey \citep{Croom:2009b} with the results from the SDSS \citep{Richards:2006a} at the bright end. They found their data to be in good overall  agreement with a double power law LF and use a luminosity and density evolution model that also included evolution in the bright-end slope (LEDE).

As a second reference we considered the AGNLF by \citet{Bongiorno:2007} (dotted line), derived from a combination of the SDSS at the bright end and the faint type~1 AGN sample from the VVDS \citep{Gavignaud:2006}. This sample is noteworthy in that it is certainly the most complete set of low-luminosity AGN at substantial $z$, as it is purely flux limited and not affected by any colour or morphological preselection. As seen in Fig.~\ref{fig:compqlf15}, the two $z=1.5$ luminosity functions are highly consistent with each other.

We then formally extrapolated these AGNLF models to $z=0$, shown as thin dotted line for the luminosity dependent density evolution (LDDE) model by \citet{Bongiorno:2007}, as thin dashed line for the LEDE model and as dashed dotted line for the LDDE model by \citet{Croom:2009}. All three extrapolations are clearly not in agreement with the local AGNLF, in several aspects: They all overpredict the space density of bright AGN. The LDDE model by \citet{Bongiorno:2007} and the LEDE model by \citet{Croom:2009} also underpredict the space density of low-luminosity AGN, while the LDDE model by \citet{Croom:2009} is roughly consistent with our data at the faint end.
These mismatches demonstrate that the local AGNLF provides indeed additional and independent constraints for the evolution of the AGN luminosity function. The above evolution models have been derived from higher redshift data which they fit very well; evidently, these models may not be extrapolated outside the range where they were observationally established. Thus, a direct determination of the AGNLF at all $z$, including $z\approx 0$, is essential.

Comparing the $z=1.5$ \citet{Bongiorno:2007} or \citet{Croom:2009} and the local AGNLF reveals a striking change in the shape of the luminosity function: The pronounced break visible at higher $z$ is almost absent in the local LF. In terms of space densities, this implies that while high-luminosity AGN were \emph{much} more frequent at high redshifts, AGN with nuclear luminosities around $M_B \sim -19$ are as  common in the local universe as they were at high $z$. For somewhat fainter AGN this relation might even be reversed, although there are too many uncertainties to make such a claim. 

Such a `downsizing' behaviour of the AGN population was first detected through X-ray surveys \citep{Ueda:2003,Hasinger:2005} and recently also in optical surveys \citep{Croom:2009}. Here we strongly confirm the presence of this `downsizing' behaviour in the optical AGNLF which becomes increasingly prominent at low redshifts. We especially see clear evidence for evolution in the faint-end slope of the AGNLF. Note that this conclusion does not hinge on our extrapolation of the H$\alpha$-$M_B$ relation, as the crossing of the $z=1.5$ and $z=0$ LFs occurs at luminosities still covered by the HES.

\section{Conclusions}

We have presented a new determination of the local ($z\approx 0$) luminosity function of broad-line Active Galactic Nuclei. Our sample was drawn from the Hamburg/ESO Survey and contains 329 quasars and Seyfert~1 galaxies with $z<0.3$, selected from surveying almost 7000~deg$^2$ in the southern sky. As a central feature, our broad-band magnitudes were measured in the survey data with a point-source matching approach, strongly reducing the contribution of host galaxy flux to the inferred AGN luminosity. Compared to our previous work we have not only substantially increased the statistical basis, but also added a number of methodical improvements. 

In the construction of the broad band ($B_J$) luminosity function, we now included the effects of differential number density evolution within our narrow redshift range, $0\la z < 0.3$. Since the most luminous AGN tend to be located near the outer edge of that range,  ignoring evolution makes the luminosity function appear slightly too shallow. We find that the evolution-corrected local luminosity function within $-19 \la M_{B_J} \la -26$ is well-described by a single power law of slope $\alpha = -2.6$, still significantly shallower than the $z=0$ extrapolation of the AGNLF measured at higher redshifts.

As a second and independent measure of AGN power we investigated the distribution of Balmer emission line luminosities, in particular the broad H$\alpha$ and H$\beta$ lines. These lines can be detected and accurately isolated in optical spectra even of low-luminosity AGN where the host galaxy is bright compared to the nucleus. We found a very tight correlation between H$\alpha$ luminosities and broad band absolute magnitudes $M_{B_J}$ over the entire luminosity range of our sample, confirming that host galaxy contamination to the $M_{B_J}$ magnitudes is unimportant.

We constructed the broad emission line luminosity functions for H$\alpha$ and H$\beta$, and found them to agree well with the broad band LF. In particular, there is again no trace of significant curvature over the covered luminosity range, and a single power law is still sufficient to describe the shape of the LF.  

We found excellent consistency between our data and the H$\alpha$ emission line luminosity function of low-luminosity AGN determined from the SDSS by \citep{Hao:2005}. While the two datasets are complementary in luminosity coverage, our low-luminosity end overlaps very well with their high-luminosity end. The SDSS data seamlessly continue the rise of the LF towards the domain of low-luminosity Seyferts. The comparison with SDSS also delineates clearly that below $L(\mathrm{H}\alpha) \sim 10^{42}$~erg~s$^{-1}$ (or $M_{B_J}\sim -19$), the HES sample becomes heavily incomplete; this we suspected already from the shape of the HES luminosity function alone. 

We combined the HES and SDSS results into a single $z=0$ AGN luminosity function covering more than 4 orders of magnitude in luminosity. This remedies a long-standing shortcoming of AGN demographics: Despite the heroic survey efforts, there was no really well-determined \emph{local} luminosity function that could serve as anchor for a global take on AGN number density evolution.
The combined local AGNLF is still amazingly close to a single power law, but it definitely shows curvature. A good description is provided by a double power law with slopes $\alpha = -2.0$ and $\beta = -2.8$.

Comparing the combined local AGNLF with determinations at higher redshifts, we find strong evidence for luminosity-dependent evolution, in the sense that weak AGN experience a much weaker number density decline, or no decline at all, than powerful quasars. This behaviour is well established from X-ray surveys, where a systematic shift of the peak in comoving space density with luminosity is observed \citep[e.g.][]{Ueda:2003,Hasinger:2005}. Known as `AGN downsizing', it is presumably related to the anti-hierarchical mass dependence of black hole growth \citep[e.g.][]{Heckman:2004,Merloni:2004,Marconi:2004,Shankar:2004,Merloni:2008,Shankar:2007}. 
The steepening of the faint end slope towards low redshift can be understood in this scenario by the change of quasar lifetime with peak luminosity, and hence black hole mass. The more massive black hole AGN die more quickly than lower mass black hole AGN and are therefore not observable in their decaying stage of their light curve, whereas lower black hole mass AGN are observable in their less luminous stage and contribute significantly to the faint end of the LF \citep{Hopkins:2006,Gavignaud:2008}. Thus the faint end of the luminosity function should consist of a mixture of low mass black holes accreting at a high rate and higher mass black holes with low accretion rates. Investigating this question will be the subject of a forthcoming paper.

\begin{acknowledgements}
We acknowledge support by the Deutsche Forschungsgemeinschaft under its priority programme SPP1177, grants Wi~1369/22-1 (LW and BH) and Wi~1369/23-1 (LW and AS).
\end{acknowledgements}

\end{document}